\documentclass[aip,amsmath,amssymb,reprint,]{revtex4-1}

\usepackage{graphicx}
\usepackage{dcolumn}
\usepackage{bm}
\usepackage[mathlines]{lineno}

\usepackage[utf8]{inputenc}
\usepackage[T1]{fontenc}
\usepackage{mathptmx}
\usepackage{etoolbox,color}

\makeatletter
\def\@email#1#2{%
 \endgroup
 \patchcmd{\titleblock@produce}
  {\frontmatter@RRAPformat}
  {\frontmatter@RRAPformat{\produce@RRAP{*#1\href{mailto:#2}{#2}}}\frontmatter@RRAPformat}
  {}{}
}%
\makeatother
\begin{document}

\preprint{AIP/123-QED}

\title{A novel concept of fractal dimension in deterministic and stochastic Lorenz-63 systems}

\author{T. Alberti}
 \affiliation{INAF-Istituto di Astrofisica e Planetologia Spaziali, via del Fosso del Cavaliere 100, 00133 Roma, Italy}
 \email{tommaso.alberti@inaf.it}
 
\author{D. Faranda}
\affiliation{Laboratoire des Sciences du Climat et de l’Environnement, CEA Saclay l’Orme des Merisiers, UMR 8212 CEA-CNRS-UVSQ, Université Paris-Saclay \& IPSL, 91191, Gif-sur-Yvette, France}
\affiliation{London Mathematical Laboratory, 8 Margravine Gardens, London, W6 8RH, UK}
\affiliation{LMD/IPSL, Ecole Normale Superieure, PSL research University, 75005, Paris, France}

\author{V. Lucarini}
\affiliation{Department of Mathematics and Statistics, University of Reading, Reading, UK}
\affiliation{Centre for the Mathematics of Planet Earth, University of Reading, RG6 6AX, Reading, UK}

\author{R.~V. Donner}
\affiliation{Department of Water, Environment, Construction and Safety, Magdeburg–Stendal University of Applied Sciences, Breitscheidstraße 2, 39114 Magdeburg, Germany}
\affiliation{Research Department I –- Earth System Analysis, Potsdam Institute for Climate Impact Research (PIK) -– Member of the Leibniz Association, Telegrafenberg A31, 14473 Potsdam, Germany}

\author{B. Dubrulle}
\affiliation{SPEC, CEA, CNRS, Université Paris-Saclay, F-91191 CEA Saclay, Gif-sur-Yvette, France}

\author{F. Daviaud}
\affiliation{CEA, IRAMIS, SPEC, CNRS URA 2464, SPHYNX, 91191 Gif-sur-Yvette, France}

\date{\today}

\begin{abstract}
Many natural systems show emergent phenomena at different scales, leading to scaling regimes with signatures of chaos at large scales and an apparently random behavior at small scales. These features are usually investigated quantitatively by studying the properties of the underlying attractor, the compact object asymptotically hosting the trajectories of the system with their invariant density in the phase-space. This multi-scale nature of natural systems makes it practically impossible to get a clear picture of the attracting set as it spans over a wide range of spatial scales and may even change in time due to non-stationary forcing. 
Here we combine an adaptive decomposition method with extreme value theory to study the properties of the instantaneous scale-dependent dimension, which has been recently introduced to characterize such temporal and spatial scale-dependent attractors in turbulence and astrophysics. To provide a quantitative analysis of the properties of this metric, we test it on the well-known low-dimensional deterministic Lorenz-63 system perturbed with additive or multiplicative noise. We demonstrate that the properties of the invariant set depend on the scale we are focusing on and that the scale-dependent dimensions can discriminate between additive and multiplicative noise, despite the fact that the two cases exhibit very similar stochastic attractors at large scales. The proposed formalism can be generally helpful to investigate the role of multi-scale fluctuations within complex systems, allowing us to deal with the problem of characterizing the role of stochastic fluctuations across a wide range of physical systems.
\end{abstract}

\maketitle

\begin{quotation}
The collective dynamics of natural systems is the result of the dynamics of their single components, often operating on multiple spatio-temporal scales and sometimes related to intrinsic and extrinsic factors. These multiple components reflect in scaling laws, unpredictable vs. deterministic behavior, bifurcations between different regimes, and strange attractors. Here, we propose a novel concept of fractal dimension in deterministic and stochastic Lorenz-63 systems to provide a better characterization of the geometrical features of attractors at different scales.
\end{quotation}

\section{Introduction}

Since their first description by E. N. Lorenz in 1963~\citep{Lorenz63}, the existence and properties of strange attractors have been frequently discussed in the context of such diverse fields as the atmosphere~\citep{Faranda19}, climate~\citep{Nicolis84,Ghil20}, biology~\citep{Nikolov14}, and ecology~\citep{Schaffer85}, to mention only a few examples. The concept of strange attractors is strictly related to that of dissipative dynamical systems with sensitive dependence on the initial conditions. Being revolutionary at the time of its invention, it has been attracting a lot of attention, especially in the context of developing measures to quantify the geometric and dynamical properties of attractors~\citep{Grassberger83a} and in revising some earlier concepts on the forecast horizon of physical systems~\citep{Kolmogorov59}. A one-parametric family of measures, the so-called generalized fractal dimensions $D_q$, has been proposed based on a coarse-grained invariant measure linking the geometric properties of the phase-space trajectories to the statistics of the dynamical scaling properties~\citep{Hentschel83}. These measures provided new insights not only in the field of dynamical system theory (where they have been developed)~\citep{Kaplan79} but also into different more applied fields like fluid and magneto-hydrodynamic turbulence~\citep{Benzi84,Macek05} and others~\citep{Cencini13}. 

One of the peculiar aspects of physical systems is their variability over a wide range of scales, arising from both intrinsic interactions between characteristic variability components in one or several variables and external forcings, differently affecting the specific properties of the whole system at different scales~\citep{Ghil20}. Recently, \citet{Alberti20} proposed a method to investigate how scale-dependency affects the phase-space properties and their statistical measures. This method requires to first identify scale-dependent components contributing to the observed dynamics of a given system as a whole, which can be achieved by applying time series decompositiion techniques like empirical mode decomposition (EMD). Subsequently, quantitative scale-specific measures like generalized fractal dimensions are evaluated. The formalism resulting from the combination of those two approaches allows the introduction of {\em multi-scale measures} by computing the generalized fractal dimensions for each scale-specific component and partial sums thereof~\citep{Alberti20}. The suitability of this approach has been demonstrated for several dynamical systems of different complexity, synthetic noisy signals, and real-world time series data~\citep{Alberti20}. For systems exhibiting heterogeneous phase space structure or even nonstationarity, it would however be useful to obtain a more detailed characterization of the time (and, hence, state) dependent multi-scale dynamical characteristics. Specifically, we are  interested in measuring the instantaneous number of degrees of freedom of a dynamical system, which is closely related to its associated recurrence characteristics~\citep{Faranda12}.

Accordingly, in this work we thoroughly extend the existing formalism of multi-scale measures \cite{Alberti20} to characterize the instantaneous scale-dependent properties of strange attractors by combining time series decomposition methods with concepts from extreme value theory that are related to the instantaneous number of degrees of freedom of the observed dynamics. We introduce and demonstrate the usefulness of the resulting concept of instantaneous scale-dependent dimension, computed on the intrinsic components of a physical signal at different scales. Notably, this approach can be used as a source of local (in terms of scales) information about the properties of the phase-space geometry of the system under study. Specifically, we demonstrate the utility of our approach for the case of the well-known low-dimensional deterministically-chaotic Lorenz-63 system and two stochastic versions thereof~\citep{Chekroun11}. We focus our attention on this perturbed system because it features different large and small scale dynamical features that can be modified by changing especially the noise level and the type of noise (additive versus multiplicative). By applying the proposed formalism, we illustrate that the properties of the system's invariant set crucially depend on the scale we are focusing on and that a global analysis will essentially reveal the large scale properties, hiding much information on the interesting dynamics triggered by additional noise at small and intermediate scales, which are associated with nontrivial resonant features.

\section{Methods}

In the following section, we start by introducing the decomposition procedure and the dynamical system metrics separately, before describing our proposed formalism. For a more general purpose, we assume to have a generic $N-$dimensional system, i.e., an $N-$dimensional phase-space, with $N>1$. Thus, we describe our decomposition procedure in a general multivariate framework. For univariate data (i.e., $N=1$), we may proceed in a largely analogous way. 

\subsection{Multivariate Empirical Mode Decomposition (MEMD)}

Considering an $N-$dimensional system described via a multivariate time series signal $\Theta_\mu(t) = [\Theta_1(t), \Theta_2(t), \ldots, \Theta_N(t)]^\dagger$ (with $\dagger$ indicating the transposition operator), the Multivariate Empirical Mode Decomposition (MEMD) decomposes the data into a finite number of multivariate oscillating patterns ${\bf C}_{\mu, k}(t)$, referred to as Multivariate Intrinsic Mode Functions (MIMFs), and a monotonic residue ${\bf R}_\mu(t)$ as
\begin{equation}
    \Theta_\mu(t) = \sum_{k=1}^{n_k} {\bf C}_{\mu, k}(t) + {\bf R}_\mu(t). 
    \label{eq:memd}
\end{equation}
The decomposition basis, formed by the set of ${\bf C}_{\mu, k}(t)$, is derived via the so-called sifting process \citep{Huang98} modified for multivariate signals \citep{Rehman10}. The sifting process consists of
\begin{enumerate}
    \item identifying local extremes of $\Theta_\mu(t)$, i.e., where the $N$-variate derivative is zero;
    \item interpolating these points via cubic splines to derive the upper and lower envelopes $\mathbf{u}(t)$ and $\mathbf{l}(t)$, respectively;
    \item deriving the mean envelope $\mathbf{m}(t)$ as $\mathbf{m}(t) = \frac{\mathbf{u}(t) + \mathbf{l}(t)}{2}$;
    \item evaluating the detail $\mathbf{h}(t) = \mathbf{s}(t) - \mathbf{m}(t)$.
\end{enumerate}
These steps are iterated until the detail $\mathbf{h}(t)$ can be identified as a MIMF (also called multivariate empirical mode) \citep{Huang98,Rehman10}, i.e., it must have the same number of local extremes and zeros (or having both differing at most by one) and a zero-average mean envelope $\mathbf{m}(t)$. The full sifting process stops when no more MIMFs ${\bf C}_{\mu, k}(t)$ can be filtered out from the data. 
Each ${\bf C}_{\mu, k}(t)$ represents a peculiar dynamical component intrinsic to the system that typically evolves on an average scale 
\begin{equation}
\tau_k = \frac{1}{T} \int_{0}^T t' \, \langle {\bf C}_{\mu, k}(t') \rangle \, dt',
\label{eq:tau}
\end{equation}
where $T$ is the length of data and $\langle \cdots \rangle$ denotes an ensemble average over the $N-$dimensional space \citep{Alberti21}.
The MEMD allows to interpret $\Theta_\mu(t)$ as a collection of scale-dependent multivariate fluctuations contributing to the collective properties of the whole system. Indeed, each MIMF can be seen as representative of fluctuations at a typical scale that is the average of the instantaneous scales (i.e., the inverse instantaneous frequencies) derived from a given mode via the Hilbert transform \citep{Alberti21}. The MEMD, due to its adaptive methodology, relieves some a priori mathematical constraints of fixed-basis decomposition methods and extracts a limited number of intrinsic components that can be visually inspected. Usual decomposition methods, like Fourier or wavelet analysis, commonly return a large number of components and/or need to project our data on a pre-defined decomposition basis. Moreover, at least classical Fourier transform based methods also require that our data satisfy a stationarity condition. In this regard, we do not question the appropriateness of the aforementioned more traditional conventional analysis techniques, but rather acknowledge that they (as well as any other approaches) have intrinsic limitations in what we can learn from their application.

\subsection{Instantaneous dimension}

Given the $N-$dimensional system described via the multivariate trajectory $\Theta_\mu(t)$, its dynamical properties can be investigated by combining the concept of recurrences in phase-space and extreme value theory~\citep{Lucarini12}. For some (arbitrary) state of interest $\zeta$ in the associated phase-space, we first introduce the logarithmic return associated with each state on the trajectory (except for $\zeta$ itself) as
\begin{equation}
    G(\Theta_\mu(t), \zeta) = -\log \left[ \text{dist}(\Theta_\mu(t), \zeta) \right]
\end{equation}
where $\text{dist}(\bullet)$ is a distance between two state vectors in phase-space, commonly the Euclidean one. By shortening the notation, we obtain a time series of logarithmic returns $g(t) = G(\Theta_\mu(t), \zeta)$ that takes larger values whenever $\Theta_\mu(t)$ is close to $\zeta$. If we now define a threshold $s(q)$ as the $q$-th empirical quantile of $g(t)$, we can introduce the exceedances $u(\zeta) \doteq \{t \, | \, g(t) > s(q) \}$, i.e., the recurrences to the reference state in the context first introduced by Poincaré. According to the Freitas-Freitas-Todd theorem the cumulative probability distribution $F(u, \zeta)$ converges to the exponential member of the Generalised Pareto Distribution (GPD), i.e.,
\begin{equation}
    F(u, \zeta) \simeq \exp \left[ -\frac{u(\zeta)}{\varsigma(\zeta)} \right]. 
\end{equation}
The GPD parameter $\varsigma$ depends on the dynamical state $\zeta$ and can be used to introduce the concept of an instantaneous dimension $d$ simply defined as $d(\zeta) = \varsigma(\zeta)^{-1}$. Although it could merely be associated to a fitting parameter, it has a clear physical meaning: $d$ is a proxy of the active number of degrees of freedom around each state $\zeta$ in the phase-space.

\section{Instantaneous scale-dependent dimension}

The instantaneous dimension $d$ introduced above provides a local view of the properties of phase-space trajectories, i.e., allows obtaining information for each point of the global structure of attractors. Nevertheless, multi-scale systems could have a scale-dependent phase-space structure~\citep{Alberti20} such that we can distinguish between features that emerge at different scales. To provide a scale-dependent instantaneous view of a given system we have to combine a decomposition method, as the MEMD, and the extreme value theory applied to inter-state distances in phase space. 

Given again an $N-$dimensional system described via $\Theta_\mu(t)$ with a multi-scale nature, i.e., being characterized by processes occurring over a wide range of scales, we can write
\begin{equation}
    \Theta_\mu(t) = \langle \Theta_\mu(t) \rangle + \sum_\tau \delta \Theta_\mu^{(\tau)}(t)
    \label{eq:avg}
\end{equation}
where $\langle \Theta_\mu(t) \rangle$ is a steady-state time-average value and $\delta \Theta_\mu^{(\tau)}(t)$ is a component of the system operating at a mean scale $\tau$. It is easy to note the analogy between Eq.~(\ref{eq:avg}) and Eq.~(\ref{eq:memd}) via the correspondence ${\bf C}_{\mu, k}(t) \leftrightarrow \delta \Theta_\mu^{(\tau)}(t)$ and ${\bf R}_\mu(t) \leftrightarrow \langle \Theta_\mu(t) \rangle$. This means that for each scale $\tau$ we can identify the corresponding invariant set $\mathbb{M}_{\tau}$ as the manifold obtained via the reconstruction of MIMFs with scales $\tau_\star < \tau$, i.e., 
\begin{equation}
    \Theta_\mu^{\tau}(t) = \sum_{k^\star=1}^{k} {\bf C}_{\mu, k^\star}(t).
    \label{eq:Thetastar}
\end{equation}
Then, for each scale $\tau \in [\tau_1, \tau_{n_k}]$, i.e., for each $k \in [1, n_k]$, given a trajectory $\Theta_\mu^{\tau}(t)$ and a state of interest $\zeta_\tau$, the cumulative probability of logarithmic returns in the neighborhood of $\zeta_\tau$ follows a GPD as
\begin{equation}
    F(u_\tau, \zeta_\tau) \simeq \exp \left[ -\frac{u_\tau(\zeta_\tau)}{\varsigma_\tau(\zeta_\tau)} \right]. 
\end{equation}
Thus, we can introduce $D(t, \tau) = \varsigma_\tau(\zeta_\tau)^{-1}$, representing the number of active degrees of freedom around each state $\zeta_\tau$. In this way, we exploit the properties of MEMD in deriving local (in terms of scale) components embedded into a given system and the instantaneous (in terms of time) properties of the extreme value theory based metric to derive the instantaneous scale-dependent metric $D(t, \tau)$. 

Summarizing, our procedure consists of the following steps:
\begin{enumerate}
 \item extract intrinsic components ${\bf C}_{\mu, k}(t)$ and their mean scales $\tau_k$ from $\Theta_\mu(t)$ by using the MEMD;
 \item evaluate partial sums of Eq. (\ref{eq:memd}) at different scales 
 \begin{equation}
  \Theta_\mu^{\tau}(t) = \sum_{k^\star=1}^{k} {\bf C}_{\mu, k^\star}(t)
  \label{eq:rec}
 \end{equation}
with $k^* = 1, \dots, n_k$ (by construction, MIMFs are ordered with increasing scales, i.e., $\tau_{k^\prime} < \tau_{k^{\prime\prime}}$ if $k^\prime < k^{\prime\prime}$); 
 \item for each scale $\tau_k$ (i.e., for each $k$) evaluate $D(t, \tau_k)$.
\end{enumerate}
Our procedure is, by construction, complete, since when $k \to n_k$ then $D(t, \tau) = d(t)$, with $d(t)$ being the instantaneous fractal dimension of the full system~\citep{Lucarini12,Faranda12,Faranda19}. 

In the remainder of this work, we will discuss some examples to highlight the potential of our framework to disentangle distinct dynamical components of different origin in a multi-scale complex system. 

\section{The Lorenz-63 model and its stochastic versions}

The Lorenz-63 system~\citep{Lorenz63}, originally developed as a simplified model for atmospheric convection, is one of the most famous and widely studied paradigmatic dissipative-chaotic dynamical systems~\citep{Ott02}, which can be written as
\begin{eqnarray}
    &&dx = s \left( -x + y \right) dt \label{eq:LorX} \\
    &&dy = \left( r x - y - x z \right) dt \label{eq:LorY} \\
    &&dz = \left( x y - b z \right) dt \label{eq:LorZ} 
\end{eqnarray}
with the parameters $\left(s, r, b \right)$ related to the Prandtl number, the Rayleigh number, and the geometry of the atmospheric convective layer. 
With the classical set of parameters $\left(s, r, b \right) = \left(10, 28, 8/3 \right)$ the system admits chaotic solutions with all initial points tending towards an invariant set, usually termed the Lorenz attractor. It is a strange attractor whose Hausdorff dimension (and all its generalizations $D_q$~\citep{Hentschel83}) take a value of $2.05 \pm 0.02$~\citep{Grassberger83a,Grassberger83b,Hentschel83}.

A simple way to investigate the role of hidden fast dynamical components is to couple deterministic equations to a "noise" mimicking the action of unknown variables. This can be also easily done for the Lorenz-63 system by rewriting the original system in terms of a set of coupled stochastic differential equations as 
\begin{eqnarray}
    &&dx = s \left( -x + y \right) dt + \sigma \, dW_t \label{eq:LorAddX} \\
    &&dy = \left( r x - y - x z \right) dt + \sigma \, dW_t \label{eq:LorAddY} \\
    &&dz = \left( x y - b z \right) dt + \sigma \, dW_t \label{eq:LorAddZ} 
\end{eqnarray}
In nonlinear deterministic systems, such additive noise can lead to non-trivial effects~\citep{Schimansky85}, including transitions between coexisting states or attractors, shifting bifurcations, or acting as an external forcing to the intrinsic variability of the system~\citep{Benzi84,Gammaitoni98}. 

More recently, another stochastic version of the Lorenz-63 system has been proposed by \citet{Chekroun11}, considering a linearly multiplicative noise term to the original system as 
\begin{eqnarray}
    &&dx = s \left( -x + y \right) dt + \sigma \, x \, dW_t \label{eq:LorMultX} \\
    &&dy = \left( r x - y - x z \right) dt + \sigma \, y \, dW_t \label{eq:LorMultY} \\
    &&dz = \left( x y - b z \right) dt + \sigma \, z \, dW_t \label{eq:LorMultZ}
\end{eqnarray}
This system provided a first example for the existence of random attractors, extending the concept of a strange attractor, still supporting nontrivial sample measures from deterministic to stochastic dynamics~\citep{Lorenz63,Chekroun11} that have been shown to be random Sinaï–Ruelle–Bowen measures~\citep{Eckmann85}. Note that in the weak-noise limit, response theory allows one to compute explicitly the change in the expectation value of the measurable observables when perturbing an underlying chaotic dynamics with stochastic terms of rather general nature \citep{Lucarini_2012jsp}.

In the following, we apply our formalism to the three different versions of the Lorenz-63 system. In the case of the stochastic models featuring multiplicative noise, we use the It\^{o} convention for the stochastic integration, $dW_t$ is a Wiener process obtained by sampling at each time step a random variable with uniform density ($W_t \sim \mathcal{N}(0,t)$) and intensity $\sigma$. The numerical simulation of Eqs.~(\ref{eq:LorX})-(\ref{eq:LorMultZ}) is obtained by using the Euler-Maruyama method with a time resolution $dt = 5 \times 10^{-3}$ over $N = 10^7$ time steps, using the classical set of parameters $\left(s, r, b \right) = \left(10, 28, 8/3 \right)$ and $\sigma = 0.4$ as in \citet{Chekroun11}.

\section{Results}

\begin{figure*}
    \centering
    \includegraphics[width=18cm]{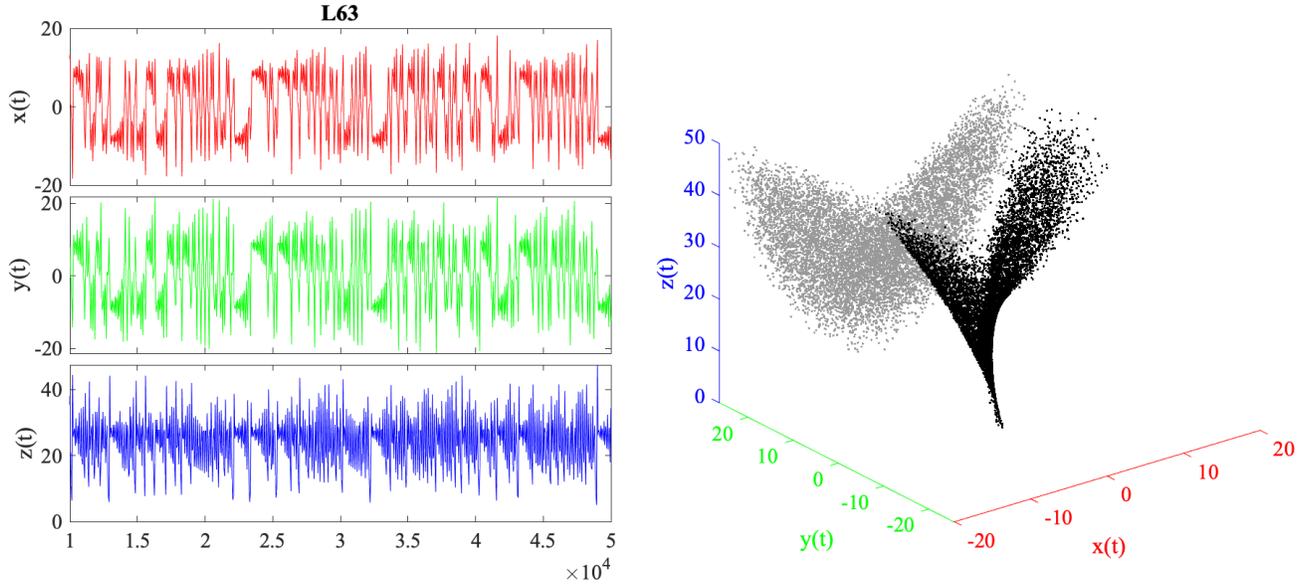}
    \caption{(Left) Zoom of the trajectory components of the deterministic Lorenz-63 system as in Eqs.~(\ref{eq:LorX})-(\ref{eq:LorZ}) (L63).
    (Right) Corresponding attractor in the 3-D phase-space (black points) and its projection in the $x-z$ plane (gray points).}
    \label{Fig1}
\end{figure*}

\begin{figure*}
    \centering
    \includegraphics[width=18cm]{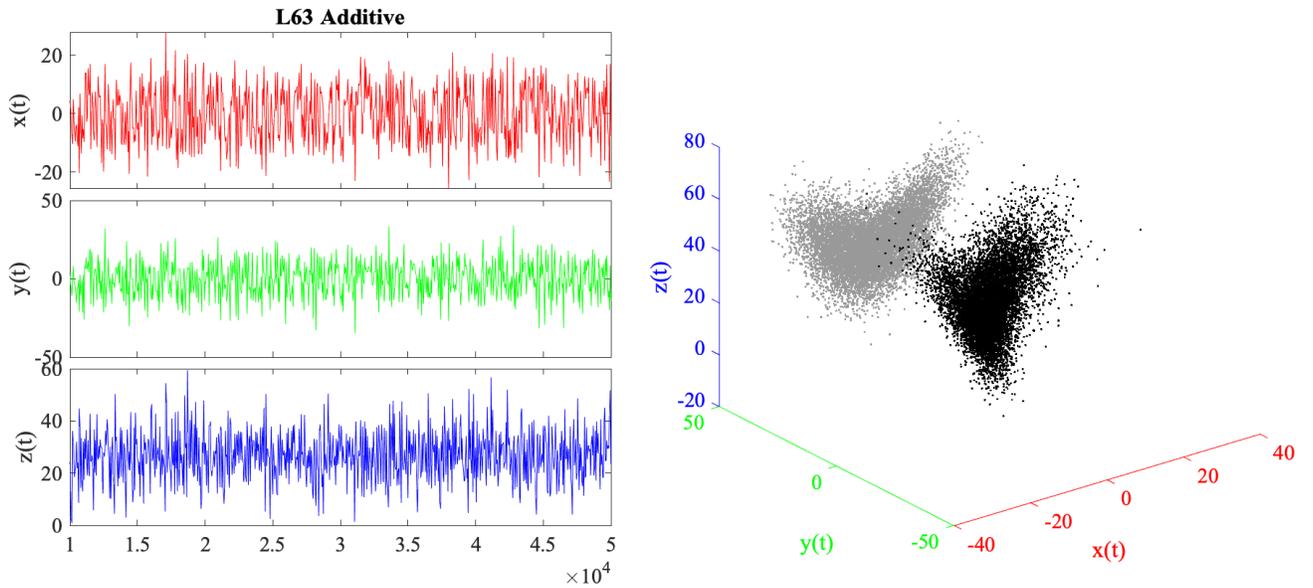}
    \caption{(Left) Zoom of the trajectory components of the additive noise model as in Eqs.~(\ref{eq:LorAddX})-(\ref{eq:LorAddZ}) (L63 Additive).
    (Right) Corresponding attractor in the 3-D phase-space. The stochastic noise term has an amplitude $\sigma = 0.4$~\citep{Chekroun11}.}
    \label{Fig2}
\end{figure*}

\begin{figure*}
    \centering
    \includegraphics[width=18cm]{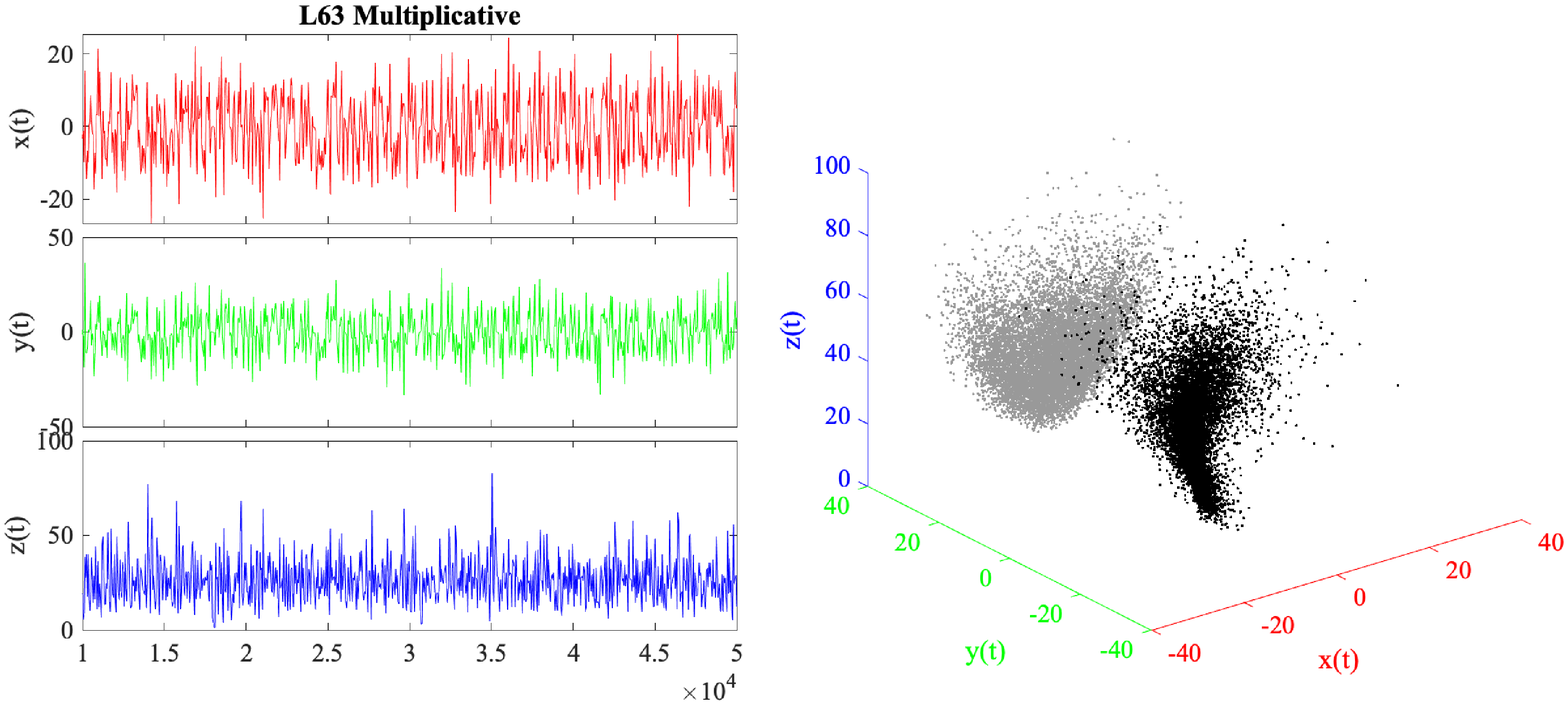}
    \caption{(Left) Zoom of the trajectory components of the multiplicative noise model as in Eqs.~(\ref{eq:LorMultX})-(\ref{eq:LorMultZ}) (L63 Multiplicative).
    (Right) Corresponding attractor in the 3-D phase-space (black points. The stochastic noise term has an amplitude $\sigma = 0.4$~\citep{Chekroun11}.}
    \label{Fig3}
\end{figure*}

Figures~\ref{Fig1}-\ref{Fig3} report the trajectories (left panels) of the three different Lorenz-63 systems (fully deterministic, Fig.~\ref{Fig1}; additive noise, Fig.~\ref{Fig2}; multiplicative noise, Fig.~\ref{Fig3}) and their corresponding attractors in the 3-D phase-space (right panels). As expected, a breakdown of the quasi-symmetric shape of the Lorenz attractor is observed when the classical Lorenz-63 system is subject to either additive or multiplicative noise. Furthermore, intermittency appears to be reduced, thus moving from a deterministic strange attractor towards a random stochastic attractor \citep{Chekroun11}.

\begin{figure*}
    \centering
    \includegraphics[width=18cm]{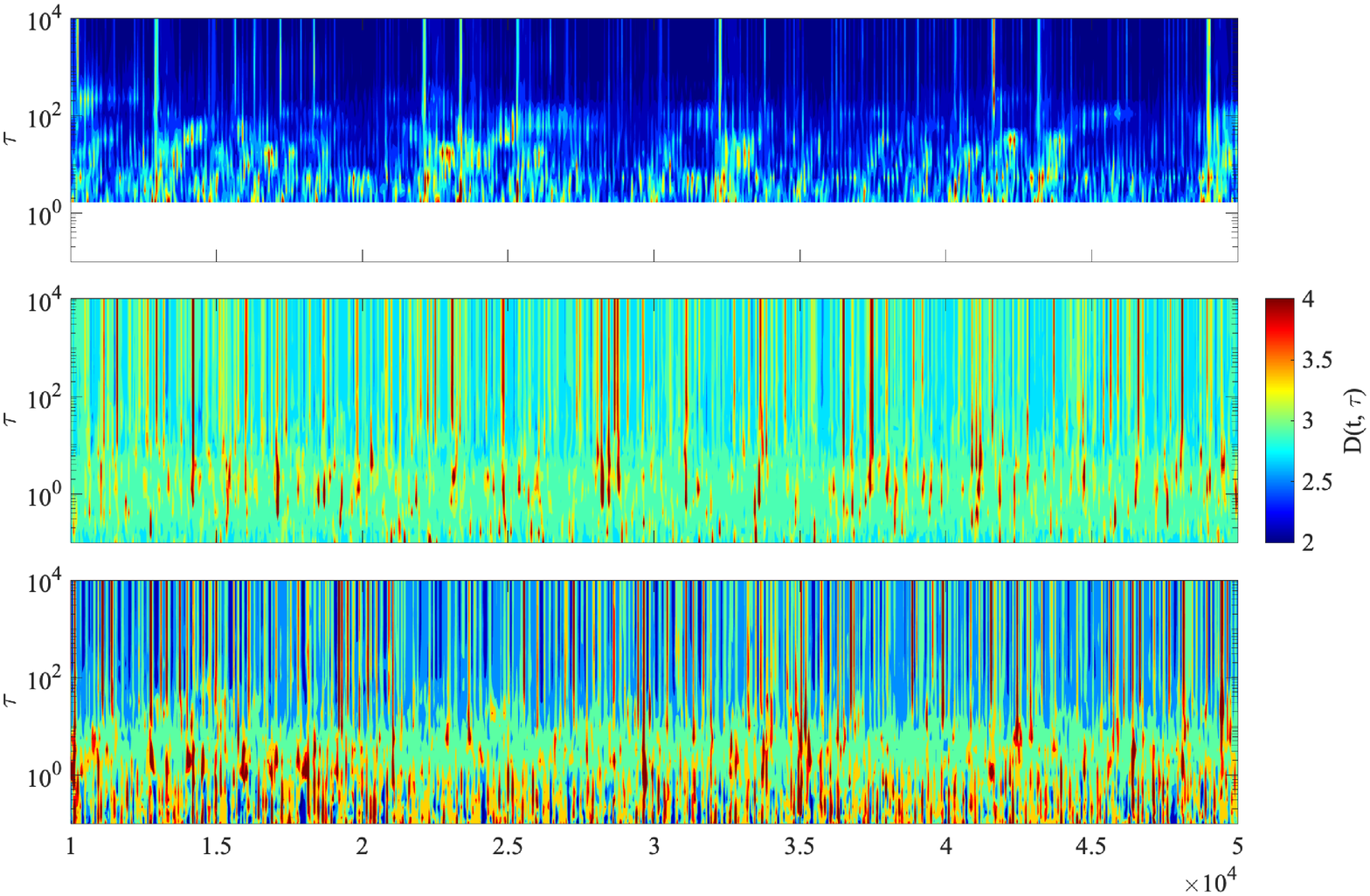}
    \caption{Behavior of the instantaneous scale-dependent dimension $D(t, \tau)$ for the deterministic Lorenz-63 system (top), its version with additive noise (middle), and the multiplicative noise case (bottom). The colormap for $D(t, \tau)$ has been saturated between 2 and 4 for a better visualization. In all three cases, an excerpt comprising 40,000 time units is shown.}
    \label{Fig4}
\end{figure*}

To further highlight these differences, we applied our formalism to derive $D(t, \tau)$ for the three different systems as reported in Fig.~\ref{Fig4}.
First of all, we evidence the absence of time scales in the range $\tau \in (3,10^2)$ for the deterministic Lorenz-63 system as opposed to its stochastic versions. This is clearly a reflection of the absence of stochastic terms in the classical deterministic system. Most notably, the traditional Lorenz-63 system is characterized by instantaneous dimension values close to 2 at timescales larger than $10^2$, almost constant along the trajectory. Conversely, values fluctuating around 3 are found for the range $\tau \in [10^0, 10^2]$. 

When a stochastic term is considered, we observe values of $D(t, \tau) \gtrsim 3$ at short timescales, extending towards larger scales when sudden transitions in the trajectory take place. This excess over the topological dimension of the phase-space is related to exiting from the region near one of the two unstable fixed points, i.e., around the lobe of the attractor, reflecting the unstable nature of the two fixed points. Typically, dimensions larger than 3 imply the existence of external forcing components, increasing the active number of degrees of freedom. Here, we can interpret this increase in terms of some extra energy provided to the system by the stochastic term, acting as an additional forcing to the autonomous dynamics. This means that the noise introduces additional degrees of freedom because it adds energy to the system: the attractor can deform through scales by increasing/decreasing its dimensions depending on the instantaneous balance between the noise forcing term and the intrinsic dynamics of the Lorenz-63 system. 

The main differences between the two stochastic versions mainly emerge at short timescales, where larger dimensions are found for the multiplicative noise case as compared to the additive one. This could be explained by invoking the fact that in the multiplicative case, the stochastic term depends on the state variables of the system. However, in both the additive and the multiplicative case, within the range of scales that can be related to the stochastic term (i.e., $\tau \lesssim 10^{0}$), $D(t, \tau)$ fluctuates around 3, with some excursions to larger values. 

\begin{figure*}
    \centering
    \includegraphics[width=18cm]{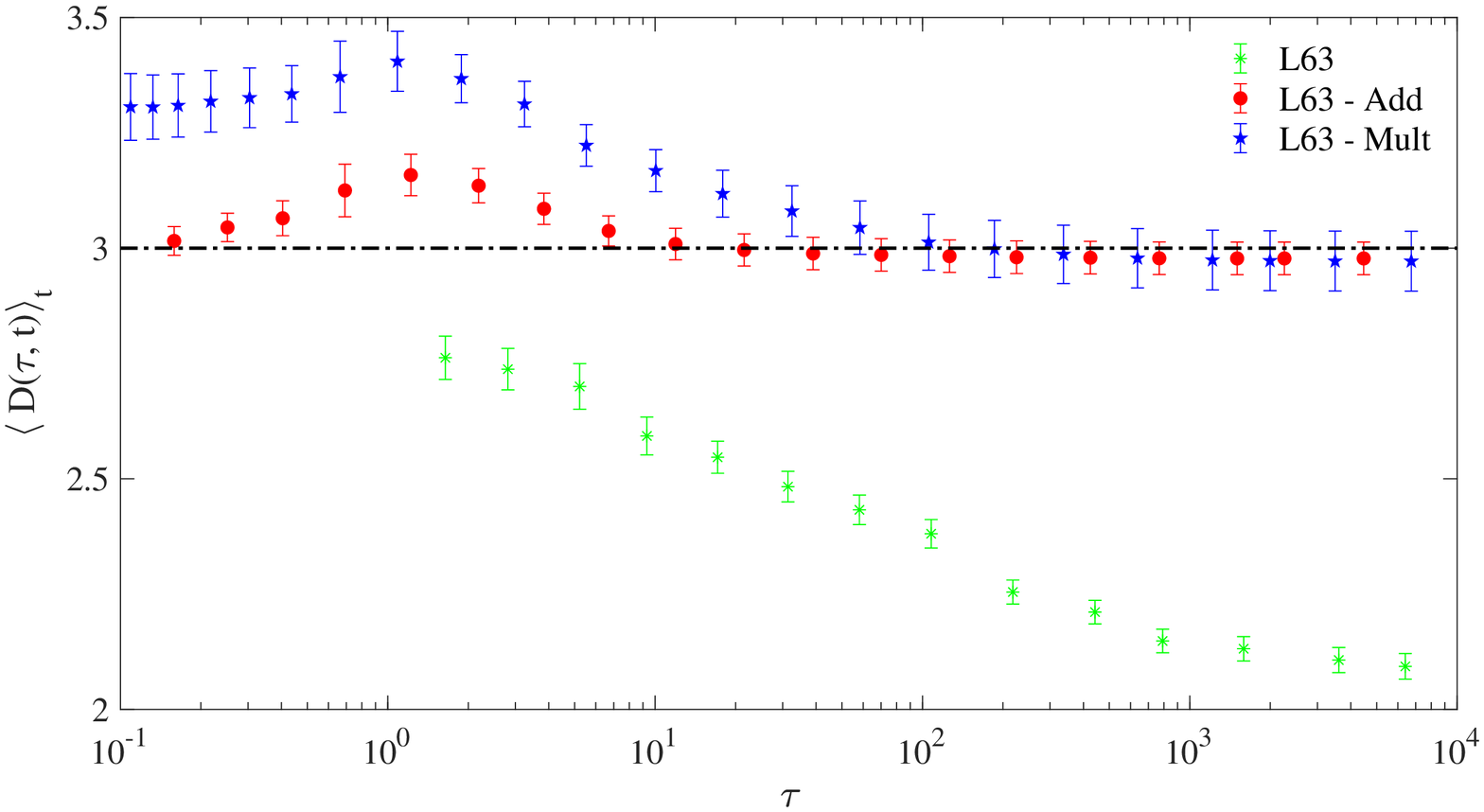}
    \caption{Average instantaneous scale-dependent dimension $\langle D \rangle$ as a function of the scale $\tau$. Green asterisks refer to the Lorenz '63 system, red circles to the additive model, and blue stars to the multiplicative case. Error bars are obtained as the standard deviations of $D(t, \tau)$ along the respective trajectory.}
    \label{Fig5}
\end{figure*}

We further evaluate the average value of our metric for the two stochastic models as compared with the deterministic Lorenz-63 system (see Fig.~\ref{Fig5}).
It is evident that larger average dimensions are found for the multiplicative noise case than for the deterministic Lorenz-63 system and the additive noise model. This reflects the effect of the stochastic term on the dynamical features of the Lorenz-63 system: it does not only act at short scales, exciting variability at additional scales with respect to the classical Lorenz-63 system, but also affects the metric at larger timescales. This can be linked to the fact that the Lorenz-63 system with non-degenerate noise has an invariant measure that is absolutely continuous with respect to Lebesgue, such that when introducing a stochastic term the dimension must converge to 3, as observed at large timescales for both stochastic models. The most interesting feature emerging for the average dimensions is that the largest value $D(t, \tau)$ for both the additive and the multiplicative case is obtained for $\tau$ of the order of the fundamental period ($\approx 1.5$ time units) of the dominating unstable periodic orbit of the deterministic system \citep{eckhardt_1994,maiocchi_2022}; see \cite{Gritsun2017} for a discussion of how unstable periodic orbits are responsible for resonant behaviour in forced systems, and \cite{Lucarini2009} for evidence of the resonant response of the Lorenz-63 system. This is likely related, as discussed before, to both the stochastic term and the intrinsic variability of the system. As expected, the average dimensions tend to saturate to those expected for the full dynamics when $\tau \to \tau_{N_k}$, being $\langle D(\tau, t)\rangle = 2.05$ for the deterministic Lorenz-63 system and $\langle D(\tau, t)\rangle = 3$ for its stochastic versions, because the invariant measure of a elliptic diffusion process has full dimension.

\begin{figure*}
    \centering
    \includegraphics[width=18cm]{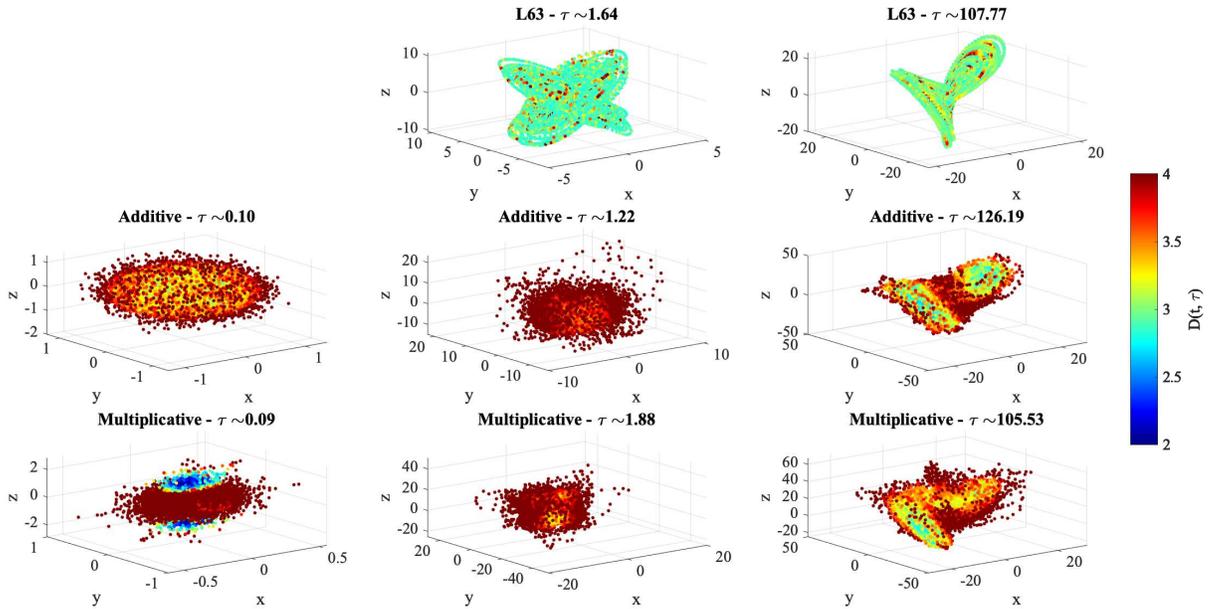}
    \caption{Three views of the Lorenz attractor at different timescales color-coded with respect the instantaneous dimensions: (top) deterministic system, (middle) additive noise and (bottom) multiplicative noise.}
    \label{Fig6}
\end{figure*}

To better highlight the scale-dependent instantaneous view of the attractor, Fig.~\ref{Fig6} reports three views of the attractor at different timescales color-coded with respect the instantaneous dimensions.
At large timescales (left panels in Fig.~\ref{Fig6}), roughly corresponding to 100 times the Lyapunov scale of the deterministic Lorenz-63 system at the considered parameter values ($\tau_L \sim 1.12$), the shape of the stochastic attractors seems to be preserved, with a qualitatively similar distribution of the values of the instantaneous fractal dimensions across the trajectory. Moreover, larger dimensions are found for the stochastic attractors than for the chaotic attractor. Averaging over time, we recover the expected values of 3 for the stochastic attractors and 2.05 for the deterministic-chaotic one. By looking at the instantaneous dimensions, we observe that $D(\tau_L, t) > 3$ at the edges of the attractor, while $D(\tau_L, t) < 3$ within the lobes. This highlights the exiting mechanisms from the region near one of the two unstable fixed points, i.e., around the lobe of the attractor, reflecting the unstable nature of the two fixed points. 

A clearly different picture can be drawn at short timescales, below the Lyapunov time (right panels in Fig.~\ref{Fig6}). Clearly, we do not have any dynamical component below the Lyapunov time for the deterministic Lorenz-63 system, indicating that variability at those fast timescales is intimately related to the stochastic forcings. While the additive case resembles a torus-like structure, the multiplicative one seems to be characterized by a saddle point-like dynamics. This seems to point towards the existence of a different fixed point structure of the origin $O = (0,0,0)$ for the two different stochastic forcings when looking at the phase-space structure at different timescales. This is due to the fact that the noise structure is different, acting as a "pure" noise term in the additive case, while playing a "forcing" role for the multiplicative one. 

Our results suggest that the stochastic term, mainly operating at short timescales, is able to change the stability of the origin, thus revealing a new structure of attractors, whose properties (i.e., its dimension) evolve in time and across scales. Thus, the attractor geometry is deformed and becomes dependent on the scale we are looking at in our system. By approaching the Lyapunov time (middle panels in Fig.~\ref{Fig6}), a restored symmetry in terms of the phase-space distribution of values of the instantaneous dimensions is observed for the stochastic attractors. Indeed, regions with low dimension are now surrounded by high-dimensional ones for both the additive and the multiplicative noise. This differs from the classical Lorenz-63 system where lower dimensions are observed with a different geometrical distribution across the phase-space, likely indicating the location of weakly repulsive low-period unstable periodic orbits~\cite{Donner2010,Donner2011b}. 

Overall, our results suggest that at those timescales where the noise terms are mainly operating, the distribution of instantaneous (local) dimensions is different from the deterministic case. Conversely, when reaching larger and larger timescales, at which the intrinsic dynamics of the Lorenz-63 system becomes significant, we observe the expected distribution of dimensions across the trajectory \citep{Faranda19}. This can be related to the existence of an invariant measure that is absolutely continuous with respect to Lebesgue, such that when introducing a stochastic term the dimension must converge to 3, as observed at large timescales. 

\section{Conclusions}

We have presented a formalism to study the behavior of chaotic or stochastic attractors as a function of the timescale, indicating that when considering different timescales the concept of a single universal attractor should be revised. Specifically, using the famous Lorenz-63 system in its standard deterministic as well as two stochastically forced versions, we have demonstrated that the attractor of this system is scale dependent. 
 
To reach this conclusion, we have extended an approach recently introduced by \citet{Alberti20} to investigate the instantaneous scale-dependent properties of attractors by combining concepts from time series decomposition methods and extreme value theory applied to recurrences in phase space. More specifically, we have used the Multivariate Empirical Mode Decomposition (MEMD) to derive intrinsic components of a given system at different timescales. Based on this decomposition, we have estimated the instantaneous scale-dependent dimensions of the system's attractor at different scales. We have show that a new structure of attractors, whose properties evolve in time, space and scale, is discovered by looking for fixed points and following their evolution from small to large scale and vice versa. Thus, the geometric structure of the attractor is gradually deformed and depends on the scale at which we are investigating the respective system.

Our formalism can be easily modified by using any alternative time series decomposition technique (like wavelet decomposition, singular spectrum analysis, or others). Our choice of the MEMD has been motivated by its empirical and adaptive nature, reducing \emph{a priori} constraints and possible artifacts of fixed-frequency/fixed-basis decomposition methods. Furthermore, the instantaneous nature (i.e., time-dependency) of the intrinsic components derived via the MEMD allows us to perform a more detailed investigation of the dynamical evolution (in time) of a system variable, better suited for evaluating instantaneous dynamical system metrics (as the dimension) than fixed-basis methods as Fourier transforms.

We are confident that the proposed formalism provides a novel way to investigate the underlying geometric (fractal) properties of physical systems at different scales during their time evolution. The concept of a scale-dependent attractor could tackle the problem of defining a more useful concept for the analysis of multiscale systems like in the case of the climate or for turbulence, which has largely remained unsolved despite numerous efforts reported in the last four decades. The corresponding prospects call for further studies to investigate these aspects in more detail, which is beyond the scope of the present paper and will be the subject of future work.

\begin{acknowledgments}
We wish to acknowledge the support by ANR TILT grant agreement no. ANR-20-CE30-0035. VL acknowledges the support received from the Horizon 2020 project TiPES (grant no. 820970) and from the EPSRC project EP/T018178/1. RVD has received funding by the German Federal Ministry for Education and Research via the JPI Climate/JPI Oceans project ROADMAP (grant no. 01LP2002B).
\end{acknowledgments}

\bibliography{Albertietal.bib}

\end{document}